\documentclass[conference]{IEEEtran}
\IEEEoverridecommandlockouts

\usepackage[utf8]{inputenc}
\usepackage[T1]{fontenc}
\usepackage{amsmath,amssymb,amsfonts}
\usepackage{graphicx}
\usepackage{booktabs}
\usepackage{textcomp}
\usepackage{placeins}
\usepackage{cite}
\IfFileExists{balance.sty}{\usepackage{balance}}{}
\usepackage[hidelinks]{hyperref}  

\begin{document}

\title{From Bytes to Behavior: Request-Level Attribution from Live GPU Hidden States}
\author{\IEEEauthorblockN{Jonathan Pan, Yidi Yuan, Xin Wei Chia, Xinhui Su, Vincent Tan, Brenda Ong}
\IEEEauthorblockA{Home Team Science and Technology Agency, Singapore \\ Jonathan\_Pan@htx.gov.sg, Yuan\_Yidi@htx.gov.sg, Chia\_Xin\_Wei@htx.gov.sg,\\Su\_Xinhui@htx.gov.sg, Vincent\_Tan@htx.gov.sg, Brenda\_Ong@htx.gov.sg}}
\maketitle

\begin{abstract}
Existing memory forensics for machine learning systems typically targets static artifacts, such as model weights or prompt logs, rather than the transient computational states of a live inference process. This gap limits an examiner's ability to analyze the specific behavior of a running large language model (LLM) service that cannot be restarted or modified. We present a method for acquiring per-layer residual-stream activations from a live GPU process using a non-modifying debugger, without requiring prior knowledge of the model's architecture parameters or weights. The approach employs a general six-stage process with platform-specific adapters, utilizing either a PyTorch allocator clock or a kernel-launch clock to identify and read tensor allocations from video memory. We validate this acquisition on 33 open-source decoder-only models across nine architecture classes, demonstrating that the recovered bytes are byte-exact and usable for downstream analysis. As a use case, we apply linear probes trained on standard framework activations to the recovered states, successfully distinguishing refusal from compliance with an AUROC of 0.88 on held-out prompts. This work establishes that live, observational access to intermediate hidden states is feasible on named inference platforms, providing a new source of evidence for examining the operational behavior of deployed AI systems.
\end{abstract}


\section{Introduction}
\label{pdb:sec:introduction}
The deployment of large language models (LLMs) has shifted toward opaque, high-throughput services where the internal mechanics of inference remain invisible to the operator. Within this black box, the transient state of the model during a specific request constitutes unique forensic evidence that static artifacts cannot provide. As a request is processed, its tokens traverse a stack of layers, and for every token and layer, the GPU holds a vector of hundreds to thousands of numbers. This vector, known as the residual stream, represents the running sum to which each layer adds its output \cite{elhage2021framework,alain2016}. Unlike weights, which define what a model is, these hidden states capture what the model did for a single request \cite{arditi2024}. However, this evidence is ephemeral; the memory holding these states is reused immediately after processing, leaving no persistent trace in standard logs or model files \cite{aip2024}.

Acquiring this transient evidence from a live GPU process presents a distinct technical challenge. Traditional memory forensics has successfully recovered keys, parameters, and even deployed models from volatile memory \cite{halderman2008lest,deepfreeze2021}, but these techniques rely on static structures or known offsets. In the context of live inference, the target is a running PyTorch process where the architecture is unknown and the memory layout is dynamic. A debugger attached to such a process cannot simply read a known address; it must identify the correct allocations from a stream of memory events without prior knowledge of the model's structure. This requires distinguishing the residual stream from other tensor allocations in real-time, a task complicated by the fact that the target process is unmodified and the acquisition must not alter the model's behavior.

This paper presents an existence proof for request-level attribution from live GPU hidden states, bounded by specific platform constraints. We demonstrate that on a live process with a non-modifying debugger attached, per-layer residual-stream rows of an unknown small LLM can be acquired without architecture parameters or weights. The method relies on a blind selector that identifies residual-stream allocations from the memory log alone, reading bytes one allocation after each tensor's birth. This approach is validated against hook references on named inference platforms. The bounds are explicit: local ptrace-capable access, a CUDA debugger, PyTorch's allocator, and decoder-only models of at most 3.4B parameters (plus one 7B model with 4-bit weights) at batch size 1. The selector required two revisions after initial failures, and the pipeline does not claim zero-knowledge acquisition or applicability to arbitrary models.

We make four contributions, each stated with its bound. First, we propose a process and adapter model comprising attach, clock, identify, read, validate, and use, where only the clock and identification rule are platform-specific. Second, we implement two adapters: an allocator clock with a blind two-step selector, which is byte-exact on 33 models across nine architecture classes, and a kernel-launch clock supporting vLLM under compilation and CUDA graphs, llama.cpp with quantized weights, and SGLang prefill, with failures reported. Third, we derive a mapped envelope with the shape of the target from the process itself, not its identity. Fourth, we demonstrate a request-level refusal call from recovered states, reporting error rates and a pre-registered test on 142 untouched prompts.

\section{Related Work}
\label{pdb:sec:related-work}
Memory forensics for machine learning systems has traditionally focused on static artifacts, such as model weights or prompt data, rather than the transient states of active inference. Early work demonstrated that memory acquired without cooperation is a legitimate source of secrets, establishing the viability of cold boot attacks \cite{halderman2008lest}. More recently, DeepFreeze recovered model parameters from edge devices, while AI Psychiatry reconstructed deployed models from CPU and GPU memory images by leveraging recovered framework objects to guide the process \cite{deepfreeze2021,aip2024}. Other studies have explored local AI inference contexts \cite{murtuza2026localai}. However, these approaches generally require knowledge of the specific framework or target static parameters, leaving the live hidden states of an ongoing inference session unaddressed.

Research on GPU data remanence and leakage has primarily targeted stale data or cross-process information, rather than the live state of a single inference request. Early forensic studies showed that data persists in CUDA global, shared, and texture memory \cite{bellekens2015}, while subsequent work identified cross-process leakage vulnerabilities \cite{cudaleaks2016}. LeftoverLocals demonstrated the recovery of LLM responses from GPU local memory left behind by another process \cite{leftoverlocals2024}, and kernel-driver forensics for NVIDIA GPUs has enabled the recovery of driver structures from RAM images \cite{bowen2024}. These methods focus on residual artifacts or inter-process boundaries, and do not capture the real-time hidden states generated during a specific request.

Work on LLM serving leaks has concentrated on prompt extraction and timing side-channels, rather than the internal hidden states. For instance, KV-cache sharing in multi-tenant environments can leak prompts through timing variations \cite{wu2025promptpeek}. Serving stacks generally fall into three categories for the purposes of this method: eager PyTorch execution; PyTorch engines utilizing block pools, paged attention, or CUDA graphs, such as vLLM which is tested here; and engines with proprietary runtimes like TensorRT-LLM, which is not tested \cite{pytorch_memory_docs}. Existing leak analyses do not examine the hidden states themselves, which remain opaque to these timing-based or prompt-focused approaches.

Linear probes and refusal directions are established techniques in interpretability, meaning the probe results presented here serve as a use case for recovered states rather than a new interpretability finding. Linear probes have long been used to read what a layer represents \cite{alain2016}. Recent work has shown that refusal in aligned chat models is mediated by a single direction in the residual stream, with per-layer probes able to decode refusal before the final layer \cite{arditi2024,conte2026,chao2024jbb,zou2023advbench,taori2023alpaca}. The refusal probe used in this paper replicates this known property on recovered states. The finding is not novel in terms of interpretability; it simply demonstrates that these established probes remain effective when applied to states extracted from live GPU memory.

\section{Threat Model}
\label{pdb:sec:threat-model}
The primary constraint governing this work is the requirement to observe a running service without altering its execution state. Standard instrumentation techniques, such as Python-level patching or `LD\_PRELOAD` injection, modify the target process or its load path, thereby changing the very evidence under examination \cite{bellekens2015,pytorch_memory_docs}. While hooks offer a simpler and faster implementation when the serving stack is cooperative, they necessitate the insertion of analyst code into the target, which is often prohibited in forensic or compliance-sensitive environments. In contrast, the method described here relies on `ptrace` access and a CUDA debugger attach to an unmodified process \cite{nvidia_cuda_gdb}. This approach preserves the integrity of the target but incurs a significant performance penalty, as the debugger halts execution at every memory allocation, resulting in slowdowns ranging from 46 to 192 times slower than native execution. The choice of this method is therefore driven by the need for hook-free observation in scenarios where the target cannot be restarted or modified, accepting the intrusion cost as a necessary trade-off for evidentiary neutrality.

The scope of what can be recovered from such live hidden states is strictly limited. The analyst possesses local debugger privilege but lacks access to the original prompt, the generated response, the specific model identity, or the causal reasoning behind any observed behavior. The evidence is stratified into two distinct levels of strength. The first is structural, comprising facts derived from the process shape, such as token count, hidden size, and layer count. The second is computational, referring to the actual intermediate states of a request. The fidelity of these computational states is verified against hook-based references only in controlled laboratory settings. Consequently, the conclusions drawn from this method are bounded by what live states can support, excluding any claims about the semantic content or intent of the underlying model interactions.

\section{Method}
\label{pdb:sec:method}
The attribution method operates as a six-stage pipeline that remains invariant across all supported platforms, with only the underlying clock mechanism and identification rule adapting to specific hardware or runtime environments. The process begins by attaching a non-modifying debugger to the live target process, followed by a clocking stage that generates an event stream indicating when each layer executes. A blind identification step then selects the residual-stream buffers using only this temporal data, without access to model architecture parameters. The system subsequently reads the raw bytes from video memory at the precise moment the clock indicates the values are final. Validation occurs once, offline, against a calibration model and a hook-based reference, ensuring that no verification logic runs during collection. Finally, the recovered states are passed to downstream analysis. Table~\ref{tbl:adapters} summarizes the specific adapter pairs tested, detailing which clock and identification rule combinations yielded byte-exact recovery across various platforms and configurations.

\begin{figure*}[!t]
\centering
\includegraphics[width=0.95\linewidth]{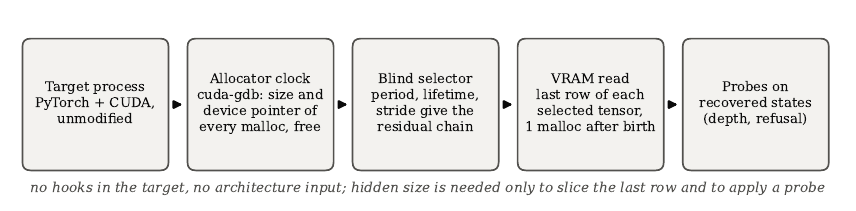}
\caption{The pipeline. A debugger attached to an unmodified PyTorch process logs every allocation; a blind selector picks the residual-stream allocations from the log alone; the bytes are read from VRAM; probes are applied to the recovered states.}
\label{fig:pipeline}
\end{figure*}

The primary clocking mechanism relies on the PyTorch caching allocator, leveraging the fact that tensor memory management is centralized in a single code path. By placing non-stopping debugger breakpoints on the allocator's internal functions, the system logs the size and device pointer of every tensor birth and death. Specifically, a breakpoint at the entry of the native caching allocator's malloc function captures the requested size and the output slot address from the argument registers, where the object pointer resides in the first register, the slot in the second, and the size in the third, consistent with the calling convention for member functions. Breakpoints at the return addresses of the two relevant call sites read the resulting device pointer, while a separate breakpoint on the free path logs each released pointer. This approach, implemented using the CUDA debugger, provides a complete record of memory lifecycle events without modifying the target application, as illustrated in the overall pipeline in Fig.~\ref{fig:pipeline}.

Identifying the residual stream from this log requires distinguishing it from other large tensors, such as key-value caches or intermediate activations. The selector operates in two steps. The primary rule identifies a hand-off chain, where a layer's output is freed immediately after the next layer's output is allocated, reflecting the fact that the output serves as the input for the subsequent operation. Formally, an allocation $a$ links to $b$ if they share the same size and $b$'s ordinal equals $a$'s ordinal plus $a$'s lifetime minus one. The lifetime of an allocation $a$, defined in (\ref{eq:lifetime}), is the count of allocations between its birth and free. The selector extracts the longest path of such links, requiring at least eight members, and trims the ends if the gap is less than half the median gap. This rule, formalized in (\ref{eq:chain}), holds for pre-normalization models. If no such hand-off chain exists, the selector falls back to a regular constant-stride chain with the longest lifetime. In this case, allocations are grouped by size and lifetime, and a group is considered regular if it contains at least eight members and at least 90 percent of consecutive gaps equal one stride of at least two. This two-step logic was revised twice following initial failures, ensuring robustness across different model architectures.

The timing of the memory read is critical to capturing the correct state. A tensor's bytes are read one allocation after its birth, because the allocation precedes the kernel that fills it. Reading at the allocation point would yield stale data, as the memory has not yet been written by the computation. Instead, the read occurs at the return of the next allocation, when the output is final and has not yet been overwritten. For the last token, the state is the final row of the tensor, so the read retrieves hidden-size times element-size bytes. This read policy was determined through measurement rather than assumption, as detailed in the negative-results table, which confirms that earlier attempts to read at different points failed to recover the correct values.

Two linear probes are trained on standard framework activations and applied to the recovered states to extract semantic information. The first probe predicts the depth, or which layer the state originates from, while the second predicts refusal, indicating whether the model is about to refuse a request. Both probes are linear models operating on standardized features, trained with a logistic or softmax loss and weight decay. The depth probe predicts the layer from a single token vector or the mean score over a tensor, using 16-token sentences from local documents for training. The refusal probe reads the residual state at the last prompt token at every layer of the chat-templated prompt. Its label is behavioral, defined as whether the greedy 32-token reply begins with a refusal phrase from a standard list. The probe score at layer $l$ for a standardized residual-stream vector $h$ is given by (\ref{eq:probe}), where $\mu$ and $\sigma$ are the per-feature mean and standard deviation on the training source, and $w$ and $b$ are the probe weights and bias.

For scenarios where the allocator clock is blind, such as when memory is managed outside the standard allocator, a second clock based on kernel launches is employed. This mechanism breaks on every kernel launch and reads the kernel's pointer arguments. Using the CUDA debugger set to break on application kernel launches, each launch stops the process, allowing the kernel name to be retrieved from the invocation column of the kernel listing. The parameter block is read through the debugger's parameter address space, starting from offset 0x160 on the sm\_75 GPU used, which is an architecture-specific value that does not require debug symbols. The block ends where a read fails. Parameter words that are 16-byte aligned and fall between $2^{32}$ and $2^{48}$ are identified as pointer candidates. For the candidates of the current and previous launch, 64 bytes of the last token's row are fingerprinted to identify the residual stream. The initial scoring method for this kernel clock failed its pre-registered criterion, prompting a corrected re-score that was pre-registered before execution. Due to the significant overhead of breaking on every kernel launch, no timing claims are made for this clocking mechanism.

\begin{table*}[!t]
\centering
\caption{Adapters built and tested: the platform-specific pair (clock, identification rule), where it worked, and where it did not. Counts are layer ends or layers recovered.}
\label{tbl:adapters}
\footnotesize
\setlength{\tabcolsep}{3pt}
\resizebox{\ifdim\width>\textwidth \textwidth\else\width\fi}{!}{%
\begin{tabular}{llll}
\toprule
\textbf{Platform} & \textbf{Clock} & \textbf{Identification} & \textbf{Outcome} \\
\midrule
HF, static KV cache & Allocator & Two-step selector & Byte-exact, 33 models; decode 6 of 7 \\
HF, default (growing) cache & Allocator & -- & No adapter: decode 0 of 8 \\
vLLM, eager & Allocator or kernel launch & Selector; or stream class & Byte-exact; stream 24/24, 22/22 \\
Any, name-free content rule & Kernel launch & Additive identity & Failed dev gate: 2 of 9 logs (eager only) \\
vLLM, default (compile + graphs) & Kernel launch & Kernel-name dataflow rule & Prefill: 31/31, 23/23, 27/27 on 3 fresh models, a duplicate-stop filter fixes long prompts; decode: 6 of 6 runs, every step correct; last layer not claimed in prefill \\
llama.cpp, CUDA backend & Kernel launch & Additive identity (prefill); norm-launch rule (decode) & Prefill byte-exact; decode 10 of 10 runs incl. last layer; quantised Q8\_0 and Q4\_K\_M 12 of 12 runs per pipeline \\
SGLang (Triton attention, radix cache off) & Kernel launch & Kernel-name rule keyed on FusedAddRMSNorm & Prefill 6 of 6 runs; decode 4 of 6 (fails on Llama-3.2-1B: wrong token count) \\
TensorRT-LLM, TGI, other GPUs & -- & -- & Untested or not testable here \\
\bottomrule
\end{tabular}}
\vspace{2pt}
\begin{minipage}{\linewidth}\footnotesize \end{minipage}
\end{table*}

\begin{equation}
\label{eq:lifetime}
\ell(a) = n_{\mathrm{free}}(a) - n_{\mathrm{alloc}}(a)
\end{equation}
where $a$ is one allocation in the log, $n_alloc(a)$ is number of mallocs logged up to and including the birth of a, within the pass, $n_free(a)$ is number of mallocs logged when a is freed.

\begin{equation}
\label{eq:chain}
a \rightarrow b \iff \mathrm{size}(a)=\mathrm{size}(b)\ \wedge\ n_{\mathrm{alloc}}(b)=n_{\mathrm{alloc}}(a)+\ell(a)-1
\end{equation}
where $a, b$ is two allocations of the same pass, $l(a)$ is lifetime of a from the previous equation, $n_alloc$ is number of mallocs logged up to and including the allocation.

\begin{equation}
\label{eq:probe}
z_l(\mathbf{h}) = \mathbf{w}_l^{\top}\,\frac{\mathbf{h}-\boldsymbol{\mu}_l}{\boldsymbol{\sigma}_l} + b_l
\end{equation}
where $h$ is residual-stream vector at layer l (a token's state), $mu, sigma$ is per-feature mean and standard deviation on the training source, $w, b$ is probe weights and bias.

\section{Experimental Setup}
\label{pdb:sec:setup}
All experiments were conducted on a single NVIDIA RTX 2080 Max-Q GPU (8 GB, sm\_75) running driver 580.178.04, CUDA 13.0, and PyTorch 2.13.0. The software stack included cuda-gdb 13.0, vLLM 0.32.0, and SGLang 0.5.23. To ensure reproducibility and prevent data leakage, development and evaluation datasets were strictly isolated. The development phase utilized 228 recovered rows from the P39 dataset, alongside earlier hook-based analyses. In contrast, Experiment A employed 142 prompts from HarmBench and XSTest that had never been used to derive rules, probes, layer selections, or thresholds \cite{chao2024jbb}. The 372 unused prompts from the local refusal-prompt file were excluded from the held-out set because they had previously been exposed to hook-based probe analysis. All frozen configuration files were hashed using SHA-256 prior to execution. For the attribution study, the model was Qwen2.5-0.5B-Instruct \cite{qwen2025}, tested on 228 chat-templated prompts ranging from 41 to 43 tokens. These prompts comprised harmful examples derived from JailbreakBench and AdvBench, as well as harmless examples derived from Alpaca \cite{chao2024jbb}. The allocator study involved 33 decoder-only models across nine architecture classes, plus five additional models with frozen rules. This study used the prompt "The capital of France is" with a batch size of one, repeated over three passes. Ground truth was established via a synchronised hook pass, kept separate from the acquisition process. Pre-registered tests were logged with hashed scripts before the run, and all failures were reported.

\section{Results}
\label{pdb:sec:results}
\subsection{Selector Robustness and Negative Results}

The two-step selector successfully identified the correct memory chain across all 33 models spanning nine architecture classes. As detailed in Table I, the selector's ordinals drove an end-to-end read session where every true layer was matched byte-exactly to reference states. This success was consistent across three passes, including one allocation after birth for hand-off chains and at the free event for the two fallback chains. The selector was never provided the layer count; it relied solely on the chain length, which included any extra members. All 33 models were tested end-to-end, with 16 operating in single-session mode. The failures observed in earlier rules are specific to certain classes rather than random noise, as evidenced by the consistent recovery of true layers in the representative models listed in the table.

Despite these successes, several alternative approaches failed, each with a distinct mechanical cause. Differential diffing of two VRAM dumps resulted in a single merged region containing all thirteen outputs, rendering it useless for isolation. Post-pass images also failed because only one of the thirteen outputs remained intact, appearing in three alternating device addresses; this indicates that depth is temporal rather than spatial. Reading at the allocation point failed for every layer, whereas reading one allocation later or at the free event succeeded. In one instance, matching ground truth by pointer alone selected an earlier temporary, briefly suggesting a model-specific read window. This episode is noted as a methodological caution against relying on pointer identity alone.

\subsection{Recovery of Refusal Probes}

A refusal probe and a difference-of-means direction computed exclusively from states recovered from VRAM reproduced the hook-based analysis. This was verified on Hugging Face and on vLLM in eager mode. The test used Qwen2.5-0.5B-Instruct, with fp32 on Hugging Face and fp16 on vLLM. The dataset comprised 228 prompts of 41, 42, and 43 tokens, processed in batches of equal length. A live single session was used, with the period found from the log and a chain of 24 selected blind. Reads of every layer's last-token vector were performed. All 228 prompts were recovered at all layers, byte-identical to the hook reference. On Hugging Face, the maximum difference was 0.0. On vLLM, the recovery was byte-exact on the 24 matched chain positions, where the chain holds the normalised layer input and includes two extra members that were not analysed. These results confirm the fidelity of the recovery process but do not constitute an interpretability claim.

\subsection{Attribution Error Rates}

An offline evidence-standard analysis was conducted on the 228 vLLM-eager recovered rows. The positive class was defined as refused, with the threshold fixed on the training source. This analysis was exploratory and not pre-registered. On JailbreakBench (71 prompts), the error rate was 15.5\% [7.0, 25.4], with a false positive rate of 13.3\% [4.4, 23.9] and a false negative rate of 19.2\% [4.5, 36.4]. This compares to a 26.8\% error rate for the dataset's own tag. On the other source (157 prompts), the error rate was 4.5\% [1.3, 7.6], with a false positive rate of 6.2\% and a false negative rate of 3.3\%, compared to an 8.3\% tag error. Table~IV summarises these results, showing that the recovered-state call beats the dataset's tag on both test sets, with error rates roughly between 5\% and 15\% depending on the source. Offline exploratory analysis (offline\_evidence\_standard.py). AUROC 0.879 and 0.986 on recovered states; hook 0.878 and 0.986.

\subsection{Pre-registered Probe Performance}

Experiment A was pre-registered, with scripts hashed before any label or probe output existed. The probe was frozen in advance and applied to states recovered from default-configuration vLLM. The test set comprised 142 prompts never used to build or tune the probe, selected from HarmBench standard behaviours and XSTest based on chat-templated length of 39 or 40 tokens. The label was the model's own greedy 32-token reply opening with a refusal phrase; 103 of 142 prompts were refused. The probe was fitted on the 228 earlier recovered rows only, with layer 21 selected by cross-validation and the threshold fixed at logit 0. All four criteria passed: token count and hidden size were derived in both buckets; 142 of 142 prompts had all 23 claimed layer ends within 2 percent; the recovered and hook AUROC differed by 0.0003; and the AUROC was 0.876 [0.808, 0.929] with a balanced accuracy of 0.777 [0.696, 0.855]. The accuracy was 0.803, compared to 0.725 for always answering refused. The false-positive rate was 0.282 [0.143, 0.436], and the false-negative rate was 0.165. The ECE was 0.160, and accuracy at coverage 0.8, 0.6, and 0.4 was 0.851, 0.918, and 0.965, respectively. Baselines included a dataset-tag rule with balanced accuracy 0.731 (AUROC 0.557) and a prompt-text classifier with AUROC 0.559. By source, XSTest unsafe had an accuracy of 0.945, HarmBench had 0.778 (18 prompts), and XSTest safe had 0.696. Table~\ref{tbl:expA} presents these metrics. The probe predicts the model's refusal behaviour, not harmfulness, and the weak points include the high false-positive rate, weak calibration, and lower accuracy on XSTest-safe and the small HarmBench subset.

\FloatBarrier
\subsection{Overhead and Length Robustness}

The acquisition process is non-modifying but not non-intrusive. Median prefill latency for Qwen2.5-0.5B-Instruct on vLLM with batch 1 was 17.5 to 17.8 ms in eager mode, 9.6 to 10.7 ms in default mode, and 17.3 ms in default mode with the debugger attached and no breakpoints. With the collector attached, the latency was 157 to 163 seconds per prompt, approximately 15,000 times the unattached default. These numbers were measured on a single GPU and should not be extrapolated to other hardware. Latency for eager mode with the collector, memory usage, decode time, and batching effects were not timed. In development-only length runs, which were not pre-registered and did not score accuracy, the unchanged rule held for token lengths of 34, 48, 100, 160, and 230. With 6 prompts each, the token count and hidden size were derived, and every claimed layer end was matched within 2 percent in 30 of 30 prompts.

\subsection{Serving Configuration and Kernel Clock}

In vLLM, the allocator clock works in eager mode but fails in the default configuration. In eager mode, the same offsets and frozen selector worked on all six models, with every layer byte-exact within four to six passes. On Qwen2.5, TinyLlama, and Llama-3.2, the chain is the output of each layer's input normalisation, not the raw stream, because vLLM keeps a (hidden, residual) pair and fuses the add into the next layer's norm. The raw stream was recovered on GPT-2 and Pythia only. In the default configuration, the allocator clock failed, with about 17 allocations per call against 164 to 192 in eager mode. No period, chain, or read was observed on Qwen2.5-0.5B and GPT-2 in-process, and on Qwen with a spawned engine process. Graph replay is the inferred, not isolated, mechanism. The kernel-launch clock still sees the kernels, with 289 stops per call in eager mode and 410.5 under graphs. A scoring defect was encountered initially, but a corrected re-score passed; both results are recorded. The kernel-name rule is keyed on the substring rms\_norm and was selected on six development logs; only the three fresh models are out of sample. The content rule found the addend row only in eager mode.

\subsection{Derived Attributes}

From the running process alone, with no model input, the acquisition derives the token count, the hidden size, and the layer count of the target. It does not identify which model it is. The token count of a forward pass and the hidden size are read from kernel-launch geometry and parameter words in a short first-pass log. The layer count is derived from the number of norm or residual-add launches in one forward pass. Table~\ref{tbl:attributes} summarises the derived attributes and the evidence for each, showing that these values were exact in the pre-registered tests across different platforms and quantisation levels.

\begin{table}[!t]
\centering
\caption{Refusal-state call from VRAM-recovered vLLM states (Qwen2.5-0.5B, 228 prompts); threshold fixed on the training source; 95\% bootstrap intervals; positive means refused.}
\label{tbl:errors}
\footnotesize
\setlength{\tabcolsep}{3pt}
\resizebox{\ifdim\width>\columnwidth \columnwidth\else\width\fi}{!}{%
\begin{tabular}{lllll}
\toprule
\textbf{Test set (n)} & \textbf{Error} & \textbf{False positive} & \textbf{False negative} & \textbf{Tag error} \\
\midrule
JailbreakBench (71) & 15.5\% [7.0, 25.4] & 13.3\% [4.4, 23.9] & 19.2\% [4.5, 36.4] & 26.8\% \\
Other source (157) & 4.5\% [1.3, 7.6] & 6.2\% [1.4, 12.1] & 3.3\% [0.0, 7.1] & 8.3\% \\
\bottomrule
\end{tabular}}
\vspace{2pt}
\end{table}

\begin{table}[!t]
\centering
\caption{Experiment A: frozen refusal probe on states recovered from default-configuration vLLM (Qwen2.5-0.5B-Instruct, 142 untouched prompts); positive means refused; 95\% bootstrap intervals.}
\label{tbl:expA}
\footnotesize
\setlength{\tabcolsep}{3pt}
\resizebox{\ifdim\width>\columnwidth \columnwidth\else\width\fi}{!}{%
\begin{tabular}{ll}
\toprule
\textbf{Measure} & \textbf{Value} \\
\midrule
Layer ends within 2\% (prompts) & 142 / 142 \\
AUROC recovered (hook) & 0.876 [0.808, 0.929] (0.876) \\
Balanced accuracy & 0.777 [0.696, 0.855] \\
Accuracy (majority baseline) & 0.803 [0.732, 0.866] (0.725) \\
False positive / false negative rate & 0.282 [0.143, 0.436] / 0.165 [0.100, 0.240] \\
Brier (base rate) / ECE & 0.167 (0.199) / 0.160 \\
Accuracy at coverage 0.8 / 0.6 / 0.4 & 0.851 / 0.918 / 0.965 \\
Dataset-label rule: balanced accuracy / AUROC & 0.731 / 0.557 \\
Prompt-text classifier: balanced accuracy / AUROC & 0.553 / 0.559 \\
\bottomrule
\end{tabular}}
\vspace{2pt}

\end{table}

\begin{table*}[!htbp]
\centering
\caption{Attributes the acquisition derives from the running process, with no model input, and how often they were exact in the pre-registered tests. Runs are model-prompt runs unless stated.}
\label{tbl:derived}
\footnotesize
\setlength{\tabcolsep}{3pt}
\resizebox{\ifdim\width>\textwidth \textwidth\else\width\fi}{!}{%
\begin{tabular}{llll}
\toprule
\textbf{Platform} & \textbf{Token count / hidden size} & \textbf{Layer count} & \textbf{Not derived} \\
\midrule
vLLM default & 6/6 prefill; 6/6 decode & 6/6 decode runs (all steps) & model identity \\
llama.cpp (fp16) & 12/12 prefill calls; 10/10 decode & 6/6 prefill pairs; 10/10 decode & model identity \\
llama.cpp (Q8\_0, Q4\_K\_M) & 12/12 prefill; 12/12 decode & 12/12 prefill; 12/12 decode & model identity \\
SGLang & 6/6 prefill; decode 4/6 (token count wrong on Llama-3.2-1B) & 6/6 prefill; 6/6 decode & model identity \\
\bottomrule
\end{tabular}}
\vspace{2pt}
\end{table*}

\section{Discussion, Limitations and Ethics}
\label{pdb:sec:discussion}
\subsection{Limitations and Scope}

The method's operational envelope is constrained by specific software dependencies and hardware configurations. It relies on PyTorch's caching allocator and build-specific offsets, meaning it is not platform-agnostic. Several serving stacks, including TensorRT-LLM and other custom runtime engines, remain untested rather than having failed. The current implementation supports models up to 3.4B parameters, with a single 7B 4-bit model tested; a 4-bit 7B mixture-of-experts model did not fit within the 8 GB memory limit. Testing was limited to batch size 1, fixed prompt lengths per session, and fp32 or bf16 precision (fp16 on vLLM). Dynamic or continuous batching and variable-length inputs disrupt the fixed pass period required for attribution. These are untested configurations, not confirmed failures, but they define the current boundary of applicability.

\subsection{Intrusiveness and Forensic Integrity}

The technique is live and intrusive, which limits its utility as a standard evidence-collection procedure. The debugger halts the process at each allocation, resulting in slowdowns between 46 and 192 times. Attaching a GPU debugger is not forensically neutral; prior work has noted that acquisition can contaminate evidence and that acquired GPU data cannot be independently verified \cite{bellekens2015}. The byte-exact comparisons reported here are possible only in a controlled laboratory setting. In a field scenario, the act of observation alters the system state, introducing a fundamental tension between measurement fidelity and operational neutrality.

\subsection{Strength of Evidence Across Levels}

The three levels of evidence do not carry equal weight. Level 1, structural attribution (token count, hidden size, layer count), is exact in every pre-registered platform test except for one instance involving SGLang decode token count on Llama-3.2-1B. Level 2, computational attribution, is the strongest technical result: the actual states of a single request are recovered with byte-exact fidelity on 33 models through the allocator clock, and located within 2\% on vLLM default for all 228 prompts. Level 3, behavioural inference, is preliminary. It relies on a refusal state with known error, tested on one model with two small held-out sources. An examiner can state structure with confidence, computation with a laboratory caveat, and behaviour only as a preliminary call with associated error rates.

\subsection{Ethical Considerations}

This paper addresses detection only. It does not describe methods for altering a model's behaviour through memory manipulation. Writing into the target's memory, steering, or removal of refusal directions are outside the scope of this work. Only open-weight models and public prompt datasets were used. Replies were truncated to 32 tokens, and only a refusal label and a 60-character prefix were retained. This minimizes the exposure of sensitive content while preserving the necessary signal for attribution.

\subsection{Generalizability and Validity Threats}

The observation principle generalizes across the architectures tested, but the fixed selector has not. It required two revisions, and the result is a map of what works rather than a universal rule. The allocator clock needs PyTorch's allocator and a build-specific offset; the kernel adapter needs a known norm kernel name and a parameter-block offset dependent on GPU architecture. The kernel-name rule held at batch 2 and on two 1.7B models but failed on one longer prompt (88 tokens), where extra norm launches shifted the layer count. This failure was traced to duplicate stop reports and fixed by a pre-filter, establishing a tested envelope of prompts up to about 100 tokens at batch 1. No platform-independent identification rule was found. Threats to validity include potential overfitting to test models after revisions, the use of keyword matching for labels, wide confidence intervals for small prompt sets (71 and 18 prompts), and testing on a single 0.5B model on one Turing GPU with batch size 1 and short prompts.

\section{Conclusion}
\label{pdb:sec:conclusion}
This work demonstrates that per-layer residual-stream rows from an unknown small language model can be extracted from a live process using a non-modifying debugger, without access to architecture parameters or weights. The approach operates on named platforms within a measured operational envelope. In our evaluation, recovered vLLM states enabled a request-level refusal classification with error rates of 5 to 15 percent on held-out sources. On a set of 142 untouched prompts, a frozen probe achieved an AUROC of 0.88, though it carried a 28 percent false-positive rate. These results indicate that the method can prioritize suspicious requests but cannot serve as a definitive conclusion. The technique reveals the model's structural shape rather than its specific identity. It incurs a cost of minutes per prompt for the target process and requires a distinct adapter for each platform.

\label{pdb:sec:--bibliography--}
\ifdefined\balance\balance\fi
\bibliographystyle{IEEEtran}
\bibliography{refs}

\end{document}